# Title: Agriculture driving male expansion in Neolithic Time

Authors: Chuan-Chao Wang[1], Yunzhi Huang[1], Shao-Qing Wen[1], Chun Chen[1,2], Li Jin[1, 3], Hui Li[1,*]

Affiliations:
1. State Key Laboratory of Genetic Engineering and MOE Key Laboratory of Contemporary Anthropology, School of Life Sciences, Fudan University, Shanghai 200433, China
2. Department of Cultural Heritage, Fudan University, Shanghai 200433, China
3. Chinese Academy of Sciences Key Laboratory of Computational Biology, CAS-MPG Partner Institute for Computational Biology, SIBS, CAS, Shanghai 200031, China
* Correspondence to: lihui.fudan@gmail.com

## Abstract

The emergence of agriculture is suggested to have driven extensive human population growths. However, genetic evidence from maternal mitochondrial genomes suggests major population expansions began before the emergence of agriculture. Therefore, role of agriculture that played in initial population expansions still remains controversial. Here, we analyzed a set of globally distributed whole Y chromosome and mitochondrial genomes of 526 male samples from 1000 Genome Project. We found that most major paternal lineage expansions coalesced in Neolithic Time. The estimated effective population sizes through time revealed strong evidence for 10- to 100-fold increase in population growth of males with the advent of agriculture. This sex-biased Neolithic expansion might result from the reduction in hunting-related mortality of males.

Agriculture first appeared in the Fertile Crescent of West Asia about 11–12 thousand years ago (kya), with domesticating few wild plant and animal species. There is also evidence of the cultivation of rice and millet in Yangtze and Yellow River Basins of China approximately 9 kya. During the subsequent times between 9 kya and 4 kya, agriculture arose independently in New Guinea highlands (9-6 kya), West Africa (5-4 kya), Central Mexico and Northern South America (5-4 kya), and Eastern USA (4-3 kya) *(1)*. Agriculture had spread to Europe, rather than independently developing there, during 9-6 kya *(2)*. Because of food production can support far higher population densities compared to hunting and foraging, major population expansions has long been suggested to has begun after the invention of agriculture *(3, 4)*. However, mitochondrial DNA (mtDNA) analysis in worldwide populations found that most major maternal lineage expansions began after Last Glacial Maximum (LGM, about 15 kya) but before the first appearance of agriculture, and the increase of population size was likely the driving forces that led to the advent of agriculture *(5, 6, 7, 8)*. There are also numerous studies on population expansion using paternal Y chromosome single nucleotide polymorphism (SNPs) in the genealogical tree and rapidly mutating short tandem repeats (STRs) *(9)*. However, nonrandom sampling of SNPs can result in an ascertainment bias, and to choose whether the evolutionary rate or the genealogical rate of STRs in Y chromosome dating is controversial, since the result can be almost three-fold difference *(9)*. With the advent of next-generation sequencing technology, entirely sequenced Y chromosomes in numerous human individuals have only recently become available *(10, 11, 12, 13, 14)*. For instance, the 1000 Genomes Project has sequenced whole Y chromosomes from more than 500 males *(14)*, which provides a wonderful chance to estimate population sizes through time from a set of globally distributed populations without ascertainment bias and helps to solve the long dispute between agriculture and initial population expansion in Neolithic Time.

We analyzed about 8.9 mega-base pairs on the unique regions of Y chromosome and whole mitochondrial genomes of 526 male individuals from three African, five European, three Asian and three American populations sequenced in the 1000 Genome Project (Table S1) *(14)*.

A maximum likelihood tree was constructed using Y chromosomal SNPs (Fig. 1, Fig. S1). Containing samples from haplogroup A, B, C, D, E, G, I, J, N, O, Q, R, and T, the tree is a good representation of geographical paternal lineages. To infer the expansion time, we calculated the date of each divergence event throughout the tree using Bayesian method *(15, 16)* with a constant mutation rate of $1\times10^{-9}$ substitution/base/year *(17)*. Most of the main haplogroups (C, I, J, O, E and R) emerged between 25-30 kya right before LGM. Almost all the primary sub-clades of main haplogroups branched off after LGM and before Neolithic Time (10-15 kya). The long internal branch lengths within those primary sub-branches indicate low population growth and frequent bottlenecks in the late Upper Paleolithic Age. However, Neolithic Time saw the great booming of new emerging lineages (Fig. 1, Fig. S1). We have also identified many star-like structures in the phylogenetic tree, such as I1a, O2b, O3a2c1 (Oα and Oβ), O3a1c1 (Oγ), R1a, R1b1a2a1, E1b1a1a1g, and E1b1a1a1f1a (Fig. 1, Fig. S1). The short internal branch lengths within the star-like clades relative to the other haplogroups are interpreted as a strong signal of recent rapid population expansions of those clades. All the star-like lineages also coalesced in Neolithic Time (5-8 kya). The Neolithic expansions for R1b, Oαβγ have also been suggested in recent published whole Y chromosome sequencing papers *(10, 13)*. Those Neolithic expanding clades comprise a large proportion of current population. For instance, R1b1a2a1 is the dominant paternal lineage of Western Europe, accounting for more than half of the population *(18)*. Oα, Oβ, and Oγ reach frequencies of 40% in Chinese *(13, 19)*. O2b comprises more than 20% of Japanese and Korean *(20)*. E1b1a1a1g and E1b1a1a1f1a account for over 80% of population in many parts of West Africa, Central Africa, East Africa as well as Southern Africa *(21)*. Thus, the time estimation of lineage coalescence showed that paternal population growth could be mainly attributed to Neolithic expansion.

We then estimated effective population size through time via coalescent Bayesian Skyline plots *(15, 16)* to infer population size changes during the LGM and the advent of agriculture (Fig. 2). The plots all show evidence of population growth and also reveal clear differences in male and female population histories. We infer rapid, roughly exponential population growth in most European and Asian maternal lineages right after the LGM (12-8 kya), followed by a long period of very slow growth (since 8-7 kya). YRI in Africa and MXL in America also show the similar pattern as European and Asian populations. However, at the paternal side, all the populations show the most pronounced expansion from 6.5 kya to 2 kya, with a 10- to 100-fold increase in population size. Except CHB and JPT had a slight growth at about 13 kya, the initial male expansion of global populations all started in Neolithic Time rather than just after LGM. The estimated timing of this male growth phase fits perfectly with archaeological evidence for the advent of agriculture in each region (Table S2). We then calculated the rate of population growth per year *(5)*. The fastest population growth for most European, Asian and MXL maternal lineages occurred about 9-11 kya (Fig.2, Table S2), corresponding to the initiation of a stable, global warming period after LGM. We also found a sharp increase in maternal population size at 33-35 kya in Asian populations, which might reflect the population explosion after initial peopling of this continent. However, the fastest growth intervals for global male populations range from 1.5-3.3 kya, which are 1-3 ky later than the advent of agriculture (Fig.2, Table S2). During this period, agriculture became mature and intensive, and can sustain a rapidly growing population. In addition, all skyline plots except Asian population (Fig.2) showed obvious bottlenecks during the agricultural transition, which might be the impact of male-dominated demic diffusion of agriculture. As in the well-known case of Europe, agricultural populations from the Middle East expanded into and throughout Europe during the Neolithic Time, and gradually replaced indigenous hunter-gatherers *(22, 23)*.

The correspondence between the coalescence age of most paternal lineages and the population growing periods observed in skyline plots suggested the initial male population expansion began within the Neolithic Time, probably due to the advent and spread of agriculture. Agriculture has provided a much more stable food supply than hunting and foraging, leading to higher population fertility and infant survival rate. More importantly, as agriculture has kept male away from dangerous hunting, the reduction in hunting-related mortality of males might contribute most to this sex-biased Neolithic expansion.

## Materials and Methods

**Populations and samples**

526 male individuals from three African, five European, three Asian and three American populations sequenced in the 1000 Genome Project were included in the current analysis (Table S1) (14). For African populations, individuals with African Ancestry in Southwest United States (ASW, 24 samples); Luhya in Webuye, Kenya (LWK, 48 samples); Yoruba in Ibadan, Nigeria (YRI, 43 samples). For European populations, Utah residents with Northern and Western European ancestry (CEU, 45 samples); Finnish in Finland (FIN, 35 samples); British in England and Scotland (GBR, 41 samples); Iberian populations in Spain (IBS, 7 samples); Toscani in Italy (TSI, 50 samples). For Asian populations, Han Chinese from Beijing Normal University, China (CHB, 44 samples); Southern Han Chinese from Hunan and Fujian Provinces, China (CHS, 50 samples); Japanese in Tokyo, Japan (JPT, 50 samples). For American populations, Colombian in Medellin, Colombia (CLM, 29 samples); Individuals with Mexican Ancestry from Los Angeles, California (MXL, 31 samples); Puerto Rican in Puerto Rico (PUR, 28 samples). The only haplogroup A individual, NA21313, is Maasai from Kinyawa, Kenya. More detailed population information could be found in the homepage of 1000 Genome Project (http://www.1000genomes.org).

**Data analysis**

About 8.9Mb sequence data on the unique regions of Y chromosome and whole mtDNA data of the 514 male individuals were extracted from the 1000 Genomes Project Phase I from publicly accessible FTP sites at the European Bioinformatics Institute (ftp://ftp.1000genomes.ebi.ac.uk/vol1/ftp/) and the National Center for Biotechnology Information (ftp://ftp-trace.ncbi.nih.gov/1000genomes/ftp/) (14).

For Y chromosome, haplogroups were classified based on ISOGG phylogenetic tree at 6 September 2013 (http://www.isogg.org/) using revised WHY.pl and AMY-tree.pl scripts (24). Y chromosomal SNPs were extracted into pseudo-sequences, and a maximum likelihood tree was calculated using PhyML (version 20120412), and bootstrap values were produced using 100 subsamplings (25). Heterozygous calls were removed in constructing the tree. We used BEAST for calculating the divergence time of each node in the phylogenetic tree (15, 16). Appropriate DNA substitution model was determined with jmodeltest-2.1.4 (26, 27) for subsequent Bayesian MCMC analysis. For Bayesian MCMC analysis, the times of each cluster were estimated using BEAST1.8.0 (15, 16). Each MCMC sample was based on a run of 30 million generations sampled every 3,000 steps with the first 3 million generations regarded as burn-in. To test the assumption of molecular clock for the tree, we used PAML package v4.4 with the GTR model (28). The null hypothesis of a molecular clock cannot be rejected (P>0.05) by comparison between the models and was therefore used for the analysis. We used the GTR model of nucleotide substitution determined with jmodeltest-2.1.4 (26, 27) with a strict clock. The single nucleotide substitution rate was set as $1\times10^{-9}$/nucleotide/year (17). Each run was analyzed using the program Tracer v1.5.0 for independence of parameter estimation and stability of MCMC chains (15, 16). Bayesian skyline plots for each population were also generated by BEAST v1.8.0 and Tracer v1.5.0, using the similar settings as above.

For mtDNA, haplogroups were classified based on mtDNA PhyloTree Build 15 at 30 September 2012 (http://www.phylotree.org/) using MitoTools (29). To test the assumption of molecular clock for the tree, we used PAML package v4.4 with the HKY+G model (28). The null hypothesis of a molecular clock cannot be rejected (P>0.05). The mtDNA coding region (positions 577–16,023) was used to generate Bayesian skyline plots using MCMC sampling in the program BEAST v1.8.0 (15, 16). Each MCMC sample was based on a run of 30 million generations sampled every 3,000 steps with the first 3 million generations regarded as burn-in. A strict clock was used and prior substitution rate was assumed to be normally distributed, with amean of $2.038 \times 10^{-8}$ subs/site/year and an SD of $2.064 \times 10^{-9}$ subs/site/year (7, 8). Each run was analysed using the program Tracer v1.5.0 for independence of parameter estimation and stability of MCMC chains (15, 16).

**Population Growth Rate Calculations**

Using the Bayesian skyline plots generated from BEAST, we calculated the rate of population growth per year. Each skyline plot consisted of 100 smoothed data points, at about 300-800 year intervals. The initial population size was set as the minimum population size during the period immediately preceding population growth. We also estimated the population growth rate for the interval in our data where growth was the fastest (Table S2). We chose the exponential growth equation for this analysis as suggest in (5): $r = \ln(N_t/N_0)/t$. Where "r" represents the population growth rate per year, "N0" is the initial population size and t the amount of time since growth began.

**Acknowledgments**

This work was supported by the National Excellent Youth Science Foundation of China (31222030), National Natural Science Foundation of China (31071098, 91131002), Shanghai Rising-Star Program (12QA1400300), Shanghai Commission of Education Research Innovation Key Project (11zz04), and Shanghai Professional Development Funding (2010001).

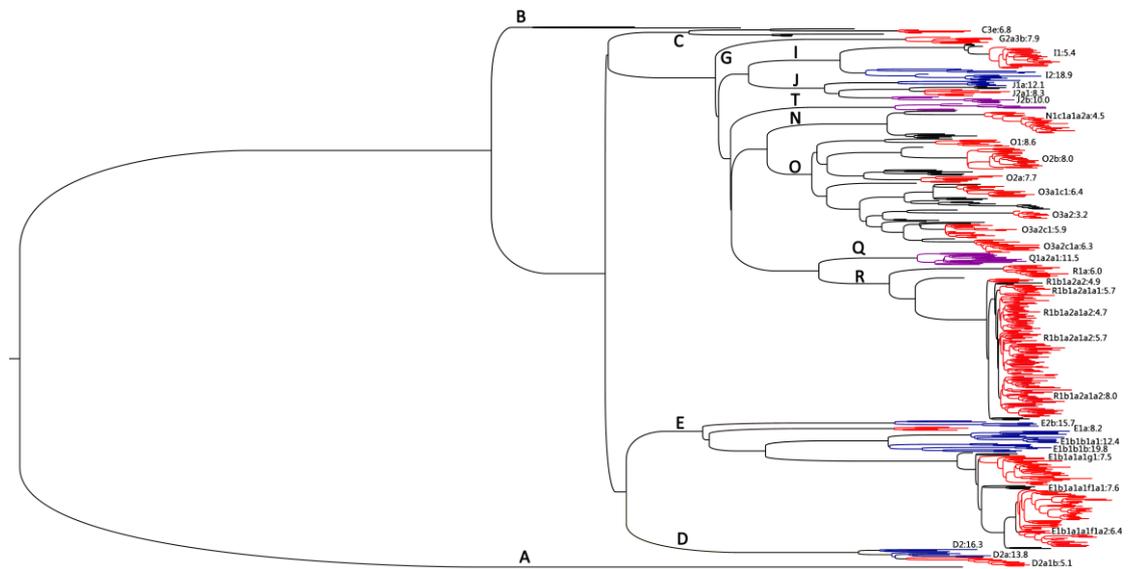

**Fig. 1.** Phylogenetic tree of human Y chromosome. This tree was constructed using 526 samples sequenced in 1000 Genomes Project. The branch lengths are proportional to the number of SNPs on the branch. Numbers in the right indicate the coalescence time (in thousand years) and 95% confidence intervals of the node are given in Fig. S1.

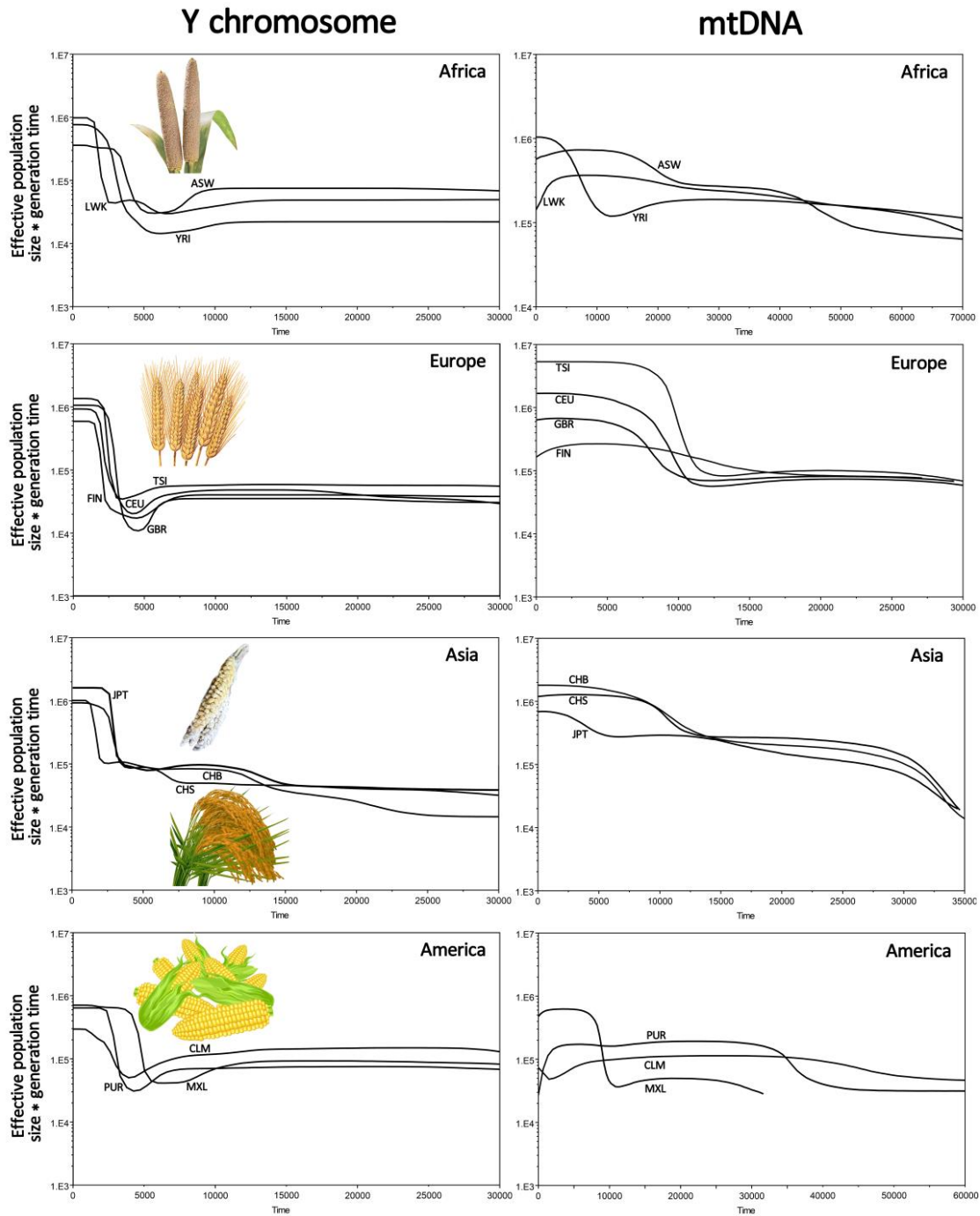

**Fig. 2.** Bayesian skyline plots of Y chromosome and mtDNA showing the effective population sizes of male and female through time. The y-axis is the product of effective size and generation time. The x-axis is the time from present in units of years. IBS was not included in this figure for its small sample size. Detailed settings refer to Supplementary Materials (Materials and Methods, Fig. S2). Population abbreviations: ASW, people with African ancestry in Southwest United States; LWK, Luhya in Webuye, Kenya; YRI, Yoruba in Ibadan, Nigeria; CEU, Utah residents with ancestry from Northern and Western Europe; FIN, Finnish in Finland; GBR, British from England and Scotland, UK; IBS, Iberian populations in Spain; TSI, Toscani in Italia; CHB, Han Chinese from Beijing Normal University, China; CHS, Han Chinese in Hunan and Fujian Provinces, China; JPT, Japanese in Tokyo, Japan; CLM, Colombians in Medellin, Colombia; MXL, people with Mexican

ancestry in Los Angeles, California; PUR, Puerto Ricans in Puerto Rico.

**Supplementary Materials**

**Figs. S1 to S2**

**Tables S1 to S2**: Table S1. Sample information and haplogroup assignment; Table S2. Population growth rates calculated from skyline plots.